\documentclass[a4paper,11pt]{article}
\usepackage{amsfonts}
\usepackage{amsmath}
\usepackage{amssymb}
\usepackage{float}
\usepackage{wasysym}

\usepackage[english]{babel}
\usepackage[latin1]{inputenc}
\usepackage{graphicx}
\usepackage{geometry}
\begin{document}

\title{Pioneer anomaly: What can we learn from LISA?}

\author{Denis Defr\`ere\thanks{Faculty of Applied Sciences, University of 
Liege, chemin des Chevreuils, 1 Bât.~B52/3 Sart Tilman, 4000 Liege, Belgium,
email: denis.defrere@skynet.be}
\and Andreas Rathke\thanks{Institute for Theoretical Physics,
University of Cologne, Zuelpicher Str. 77, 50937 Cologne, Germany,
email: ra@thp.uni-koeln.de
%EADS-Astrium GmbH, 88039 Friedrichshafen, Germany, email:
%andreas.rathke@astrium.eads.net
}}

\date{}
\maketitle

\begin{abstract}
  The Doppler tracking data from two deep-space spacecraft, Pioneer 10 and 11,
  show an anomalous blueshift, which has been dubbed the ``Pioneer anomaly''.
  The effect is most commonly interpreted as a real deceleration of the
  spacecraft --- an interpretation that faces serious challenges from
  planetary ephemerides. The Pioneer anomaly could as well indicate an unknown
  effect on the radio signal itself. Several authors have made suggestions how
  such a blueshift could be related to cosmology.  We consider this
  interpretation of the Pioneer anomaly and study the impact of an anomalous
  blueshift on the Laser Interferometer Space Antenna (LISA), a planned joint
  ESA-NASA mission aiming at the detection of gravitational waves. The
  relative frequency shift (proportional to the light travel time) for the
  LISA arm length is estimated to $10^{-16}$, which is much bigger than the
  expected amplitude of gravitational waves.  The anomalous blueshift enters
  the LISA signal in two ways, as a small term folded with the gravitational
  wave signal, and as larger term at low frequencies.  A detail analysis shows
  that both contributions remain undetectable and do not impair the
  gravitational-wave detection. This suggests that the Pioneer anomaly will
  have to be tested in the outer Solar system regardless if the effect is
  caused by an anomalous blueshift or by a real force.
\end{abstract}

\section{Introduction}

The Laser Interferometer Space Antenna (LISA) is a joint ESA-NASA mission to
be launched after 2012 that will detect gravitational waves (GWs) in a
frequency range between $10^{-4}$ and $1\,$Hz and study their sources
\cite{LISA}.  LISA will consist of three spacecraft forming a roughly
equilateral triangle of $5\times10^9$\,m baseline placed on an orbit similar
to that of the Earth.  The spacecraft will exchange phase-coherent laser
signals with each other in order to conduct picometer interferometry to
measure passing GWs through the modulation in the light travel time between
the spacecraft that the waves cause.

In this study we consider the impact of an anomalous blueshift, which is
homogeneous in the light travel time and isotropic, on LISA. The motivation to
consider such an effect comes from the Doppler tracking data of the Pioneer 10
and 11 deep space probes. Both spacecraft show a deviation between their orbit
reconstruction and their Doppler tracking signal
\cite{Anderson:1998jd,Anderson:2001sg}. The discrepancy, that has become known
as the Pioneer anomaly, can correspond either to a small constant deceleration
of the spacecraft of roughly $9\times 10^{-10}\,{\rm m/s^2}$ or to an anomalous
blueshift of the radio signal at a rate of $6 \times 10^{-9}\,{\rm Hz/s}$.
Since no unambiguous conventional mechanism, like small on-board forces, to
explain the anomaly has been identified there is a growing number of studies,
which consider an explanation in terms of a novel physical effect (see
\cite{Anderson:2001sg,Bertolami:2003ui,Izzo:2005dz} for an overview of the
theoretical models). It has been realised that it is difficult to explain the
Pioneer anomaly by a real force which satisfies all constraints from planetary
ephemerides \cite{Anderson:2001sg,Izzo:2005dz,Sanders:2005vd}. Hence an
explanation in terms of an anomalous blueshift seems particularly attractive.

In view of the increasing interest in an experimental verification of the
Pioneer anomaly \cite{Dittus:2005re} it is a logical step to consider if such
a verification might be possible with a space mission that is already planned.
Unfortunately, the current and upcoming exploration missions are hardly suited
for a verification of the Pioneer anomaly \cite{Izzo:2005dz,Nieto:2003rq}.
LISA is the first upcoming high-precision fundamental-physics mission that
might be sensitive to the anomaly. Already in an early discussion, L.~Scheffer
expressed the expectation, that the Pioneer anomaly, if not due to a
spacecraft-specific conventional reason, should be detectable in data from
LISA \cite{Scheffer:2001se}. In a proposal to ESA's `Cosmic Vision 2015-2025
Call for Themes' the question was raised again if LISA could be a suitable
testbed for a verification of the Pioneer anomaly -- in particular if the
effect were due to an anomalous blueshift \cite{RosalesProposal}.  Even more
important might be the opposite question: If the Pioneer anomaly is indeed a
novel physical effect could it impair the performance of LISA? In this case it
would be of crucial importance to ensure that the LISA science goals can be
achieved despite of the presence of the anomaly.  The present study addresses
both of these questions and comes to the conclusion that LISA is neither
sufficiently sensitive to the Pioneer anomaly to detect it nor impeded in its
mission goals by the potential presence of the anomaly.

The layout of our considerations is as follows. Section~\ref{PAandLISA} gives
an overview of the Pioneer anomaly and its possible relevance for LISA. In
Section~\ref{characteristics} we review the observational evidence for the
Pioneer anomaly, and briefly review the models, that have been put forward to
explain the anomaly.  In Section~\ref{sect:anomalous} we discuss which models
of the Pioneer anomaly are relevant for observations with LISA and derive a
first order of magnitude estimate for the maximal effect to expect on the
interferometric signal of LISA. We also find the generic response function of
LISA in the presence of an anomalous blueshift.  Section~\ref{frequency}
discusses the effect of the blueshift in the frequency domain.  The frequency
domain method has been discarded for the actual evaluation of LISA
interferometric data because time-delay interferometry (TDI) achieves a far
superior cancellation of the laser phase noise (cf. \cite{Tinto:1999yr}).
However the frequency domain method has the advantage that it gives direct
physical insight into the impact of an anomalous blueshift on the
interferometer.  Section~\ref{sect:twow} briefly reviews the structure of the
interferometric signal of LISA and its Fourier transform.  It is followed by
the analysis of the impact of the blueshift, which is split into two parts.
First Section~\ref{inside} discusses the effect of the anomalous blueshift in
the sensitivity band of LISA. Then the detectability of the blueshift at very
low frequencies outside of the sensitivity band of LISA is considered in
Section~\ref{outside}. In both cases no measurable impact of the anomaly is
found.  Section~\ref{time} reconsiders the effect of the anomaly in the
framework of TDI, the current baseline method for LISA. Section~\ref{combi}
discusses the signature of the anomaly on first generation TDI observables for
the idealised case of fixed arm length. It is found that the symmetry of TDI
observables leads to an exact cancellation of the effect of the anomalous
blueshift in the case of fixed interferometer arms.  Section~\ref{sect:TDI2}
generalises these considerations to the realistic case of moving spacecraft.
Also in this setting the effect of the anomalous blueshift would remain below
the detection threshold of LISA.  Section~\ref{conc} summarises our results
and discusses their implications for options to verify and characterise the
Pioneer anomaly.

\section{The Pioneer anomaly and LISA\label{PAandLISA} }

\subsection{The characteristics of the Pioneer 
anomaly\label{characteristics}}

The Pioneer~10 and 11 spacecraft, launched on 2 March 1972 and 5 April 1973,
respectively, were the first to explore the outer Solar system (see 
\cite{Pioneers} for an overview of the Pioneer 10 and 11 missions.). Since its
Jupiter gravity assist on 4 December 1973 Pioneer~10 is on a hyperbolic coast.
Pioneer~11 used a Saturn swingby on 1 September 1979 to reach a hyperbola, in
approximately opposite direction to Pioneer~10. Already before the swingby a
discrepancy between the Doppler signal from Pioneer~10 and its orbit
integration was observed, which was originally ascribed to fuel leaks and a
mismodelling in the Solar radiation pressure model (cf.\ \cite{Null}). This
interpretation became more and more untenable after the swingbys due to the
decrease of the Solar radiation pressure, inversely proportional to the square
of the heliocentric distance, and the quiet state of the spacecraft, with very
little thruster activity.  Moreover an anomaly of the same magnitude became
apparent in the Pioneer~11 data \cite{Nieto:2005kb}.

The anomaly on both probes has been subject to three independent analyses with
different orbit determination programs
\cite{Anderson:2001sg,Markwardt:2002ma}. The result of the investigations is
that an anomalous Doppler blueshift is present in the data from both
spacecraft of approximately $6 \times 10^{-9}\,{\rm Hz/s}$ corresponding to
an apparent deceleration of the spacecraft of approximately $9 \times
10^{-10}\,{\rm m/s}^2$.  From the Doppler data, it is not possible to
distinguish between an anomalous frequency shift of the radio signal or a real
deceleration of the spacecraft (see below).  The principle investigators of
the anomaly have conducted a thorough investigation of possible biases and
concluded that no conventional effect is likely to have caused the anomaly
\cite{Anderson:2001sg}.  Meanwhile, there exists an ample body of literature
discussing various aspects of possible systematic effects, without definitive
conclusion
\cite{Scheffer:2001se,Katz:1998ew,Murphy:1998hp,Anderson:2001ks,Mashhoon:2002fq,Anderson:2003rd}.
For various reasons all other deep-space probes have lower navigational
accuracy \cite{Anderson:2001sg,Izzo:2005dz,Nieto:2003rq}. Hence to date the
effect could not be verified with another spacecraft.

The inability to explain the anomalous acceleration of the
Pioneer spacecraft with conventional physics has contributed to the growing
discussion about its origin. The possibility that it could come from a new
physical effect is now being seriously
considered.  In particular the coincidence in magnitude of the Pioneer 
anomaly
and the Hubble acceleration has stirred the suggestion that the Pioneer
anomaly could be related to the cosmological expansion.

One of the obstacles for attempting an explanation of the Pioneer anomaly in
terms of new physics is that a modification of gravity, large enough to
explain the Pioneer anomaly, is likely to run into contradiction with the
planetary ephemerides. This is readily illustrated by adding a term
corresponding to the Pioneer anomaly to the Newtonian potential of the Sun,
\begin{equation}
V(r) = - \frac{\mu_\odot}{r} - a^* r \, ,
\end{equation}
($\mu_\odot$ is the reduced mass of the Sun, $r$ is the heliocentric distance,
$a^* \approx 9 \times 10^{-10}\,{\rm m/s^2}$ is the anomalous acceleration) and
considering the orbit of Neptune. At 30\,AU the Pioneer anomaly is visible in
the Doppler data of both Pioneer 10 and 11. The influence of an additional
radial acceleration of $9 \times 10^{-10} {\rm m/s^2}$ on Neptune is
conveniently parameterised by a change of the effective reduced Solar mass
$\mu_{\odot}$, felt by the planet (cf.\ \cite{Talmadge:qz}). The value
resulting for the anomaly, $\Delta \mu_{\odot} = a^* r_{\neptune}^2 \approx 1.4 \times
10^{-4} \, \mu_{\odot}$, is nearly two orders of magnitude beyond the current
observational constraint of $\Delta \mu_{\odot} = (-1.9 \pm 1.8) \times 10^{-6}
\, \mu_{\odot}$ \cite{Anderson:1995dw}.  Similarly the Pioneer 11 data
contradict the Uranus ephemerides by more than one order of magnitude. Thus,
the Pioneer anomaly can hardly be ascribed to a gravitational force since this
would indicate a considerable violation of the weak equivalence principle.  In
particular, planetary constraints rule out an explanation in terms of a
long-range Yukawa force \cite{Anderson:2001sg,Reynaud:2005xz}. Hence, more
subtle explanations are to be attempted.

One line of reasoning is to consider an effect on the radio signal rather than
a force on the spacecraft. Already the principle investigators have considered
several phenomenological models of accelerating time \cite{Anderson:2001sg}.
The main purpose of these models was to investigate the possibility of a
systematic drift of atomic clocks.  Most of the phenomenological models failed
the cross-check with tracking data from other spacecraft. Only a time
acceleration restricted to the signal propagation itself yielded a good fit to
all spacecraft data although this model is still statistically disfavoured to
a real deceleration of the spacecraft. The time acceleration of this model is
indistinguishable from a run-time/travel-distance dependent blueshift of the
radio signal.

To first order in $v/c$ the anomalous Doppler drift is related to the
anomalous acceleration as
\begin{equation}
\frac{1}{\nu}\frac{d\nu}{dt}=-\frac{a^*}{c} \label{bs}\, ,
\end{equation}
where $\nu$ is the emitter frequency of the signal, $v$ is the spacecraft
velocity and $c$ is the velocity of light (cf.\ \cite{Anderson:2001sg}). Note
that $a^*$ is negative since it indicates a deceleration. At first sight this
coincidence in phenomenology between an anomalous deceleration and an
anomalous blueshift is surprising. It gets explained if one considers that the
anomaly was only thoroughly investigated for the part of the Pioneer
trajectories through the outer Solar system: Here the back-reaction of the
spacecraft's orbit to a small perturbing force can be neglected and an
anomalous acceleration can be treated linearly to high accuracy
\cite{Izzo:2005dz}.\footnote{This simple observation illustrates the need for
  the analysis of the full Doppler data of Pioneer 10 and 11 because from data
  further inward in the Solar system a discrimination between a real force and
  a blueshift might be possible through the presence or absence of a change of
  the orbital parameters due to the anomaly.}

Several theoretical models have been put forward that implement an anomalous
blueshift by very distinct mechanisms
\cite{Rosales:1998mj}-\cite{Ranada:2004mf}.  The works
\cite{Rosales:1998mj,Rosales:2002ui} consider the anomaly as a kinematical
effect of the cosmological expansion.  The anomaly arises from the fact that
the coordinate system, in which local measurements are carried out, is not a
synchronous one. The studies \cite{Rosales:2004kb,Rosales:2005} consider an
adiabatic effect of the cosmic expansion on the phase of light viewed as the
phase of a quantum state.  Whereas \cite{Rosales:2004kb} considers a closed
path Berry phase, \cite{Rosales:2005} drops the closed path requirement and
considers an open path Berry phase.  In
\cite{Ranada:2002fm}-\cite{Ranada:2004mf} the anomaly arises from a time
dependence of the local metric which leads to an effective time acceleration.

All of the above models to explain the blueshift of the Radio signals
transponded by the Pioneers have to be considered as incomplete. This is most
obvious for the model of \cite{Rosales:1998mj,Rosales:2002ui}, where only a
Robertson-Walker metric is considered and the influence of the gravitational
field of the Sun is completely neglected. This seems too much of a
simplification considering the predominant opinion that the local
Schwarzschild geometry of the Solar system remains practically unaffected by
the cosmological expansion (see the contribution of C. L\"ammerzahl in this
volume). The problem is ameliorated a bit for the quantum effect considered in
\cite{Rosales:2004kb,Rosales:2005} because in this case one could argue that
the adiabatic evolution of quantum states is governed by a different metric
than the non-adiabatic dynamics of large bodies. Also the definition of the
open-path Berry phase in \cite{Rosales:2005} does not seem to be compatible
with the general discussion of the open-path Berry phase in
\cite{Pati:1998hf}. In the models of \cite{Ranada:2002fm}-\cite{Ranada:2004mf}
the embedding problem does not seem to spoil the model because both the cosmic
and the local metric are treated as perturbations of a locally flat metric and
can (at least formally) be superimposed linearly. However the model of
\cite{Ranada:2002fm,Ranada:2003ax} suffers from the introduction of two ad-hoc
coupling parameters between the electromagnetic and the gravitational field
\cite{Jafari:2004vq}. Furthermore, the models
\cite{Ranada:2002fm}-\cite{Ranada:2004mf} lack a relativistic derivation of
the background potential from the cosmological parameters.  Despite of the
deficiencies of the current models the idea that the Pioneer anomaly is caused
by a blueshift of light is attractive because it automatically satisfies all
constraints from planetary ephemerides.

\subsection{Relevance for LISA \label{sect:anomalous}}

Among the proposed explanations of the Pioneer anomaly, most would
have no significance for LISA.  For example, this is the case for
all models based on systematics generated onboard the Pioneer spacecraft.
Generally, if the anomaly corresponds to a real acceleration on
the Pioneers, the anomaly should have no influence on LISA. This
can be concluded from the fact, that the LISA orbit is practically
identical to the Earth's orbit. For the Earth itself an anomalous
acceleration of the magnitude of the Pioneers would lead to an
orbital perturbation which is beyond current observational limits
(cf. \cite{Talmadge:qz}). Hence only a considerable violation of
the weak equivalence principle (e.\,g.\ between bodies of
different mass) could result in an anomalous acceleration on LISA
but not on the Earth.
On the other hand an anomalous blueshift of light
could be highly relevant for LISA, since the mission is
supposed to detect GWs through small frequency
shifts.
The blueshift for light travelling along an arm of LISA is found
by integrating Eq.~(\ref{bs}) in time,
\begin{equation}
\frac{\Delta \nu^*}{\nu_0}=-\frac{a^*}{c}\,T \label{blue}\, ,
\end{equation}
where $T$ is the light travel time and $\nu_0$ is the laser
frequency. For the LISA values, $\nu_0 \approx 3 \times 10^{14}\, {\rm
Hz}$ and $T \approx 17\,{\rm s}$, one finds $\Delta \nu^* \simeq 1.5
\times 10^{-2}\,{\rm Hz}$.  Although the absolute blueshift is very
small compared with the nominal frequency it might nevertheless be
within the reach of LISA.  Indeed, the corresponding relative change
of the frequency is $\Delta \nu^*/\nu \simeq 10^{-16}$ and the
expected value for the weakest GWs, that will be detectable by LISA,
is about $10^{-23}$ \cite{LISA}.  The frequency shift
due to the anomaly is therefore seven orders of magnitude bigger than
the lowest signal to be detectable by LISA.  The ability to measure
the contribution of the anomalous blueshift will however depend on the
sensitivity of LISA at the frequencies where the anomaly is present.

For a comprehensive analysis of the impact of the anomalous blueshift on 
LISA
one has to take into consideration the change of the light
travel time by passing GWs. In linear order in the
GW strain $h$ the rate of change of the light travel time caused by a plane wave
is proportional to the projection of the difference of the GW
strains at the point of reception and the point of emission
onto the light travel direction \cite{Esta:1975,Hellings:1981xc},
\begin{equation}
\frac{d}{dt} \Delta T = \frac{1}{2}(1+\beta)
\big( h(t)- h[t+(1-\beta)T] \big) \, .
\label{EstaWahl}
\end{equation}
Here $T$ is the unperturbed one-way light travel time, $\Delta T$ is the
change of light travel time and $\beta$ is the cosine of the angle between
the light travel direction and the normal of the wave front of the GW. The
time $t$ is the time measured by a clock at the point of reception.

Writing Eq.~(\ref{blue}) for the modified light travel time $T + \Delta T$ 
and
using Eq.~(\ref{EstaWahl}) to express $\Delta T$ by the linear term of a
Taylor expansion, one obtains the frequency shift for the combined
effect of the anomalous blueshift and GWs up to linear order in $a^*$ and 
$h$
for a one-way signal,
\begin{equation}
\frac{\nu_1-\nu_0}{\nu_0}
= -\frac{a^*}{c}T
+ \frac{1}{2}(1+\beta) \left[1-\frac{a^*}{c}T \right]
\big( h(t)-h[t+(1-\beta)T] \big)\, ,
\label{oneway}
\end{equation}
where $\nu_1$ is the frequency
at reception. Eq.~(\ref{oneway}) is generic
for any model of a homogeneous isotropic blueshift or time-acceleration. In
particular it holds for the models considered in
%\cite{Anderson:2001sg,Rosales:1998mj,Rosales:2002ui,
%  Rosales:2004kb,Rosales:2005, Ranada:2003ax,Ranada:2004mf}.
\cite{Anderson:2001sg}-{Ranada:2004mf}.

Depending on the model there might arise one subtlety, which has not been
addressed up to now. In the same way as the electromagnetic waves are
blueshifted an analogous blueshift might arise for the GWs.  For example, this
is the case in the models of \cite{Ranada:2002fm}-\cite{Ranada:2004mf}, where
the anomalous blueshift originates from a time-dependent term in the $g_{00}$
component of the metric caused by a homogeneous cosmological background
potential. This additional term leads to a modified dispersion relation for
all types of waves. The anomalous blueshift of GWs could be investigated by a
parameter estimation of the dispersion relation via matched filtering of GW
signals detected by LISA. The method would be analogous to the search for a
graviton mass in GW signals (cf.\ 
\cite{Will:1997bb,Will:2004xi,Berti:2004bd}). In the present study we restrict
ourselves to the possible manifestations of the Pioneer anomaly in
electromagnetic waves because the occurrence of a blueshift of GWs is model
dependent and hence would hardly allow a generic statement about the LISA's
capability to verify the Pioneer anomaly.

Rather than the one-way response function of Eq.~(\ref{oneway}), the two-way
response function of a signal transponded back to its emitter is the 
relevant
observable for LISA. It is found analogous to Eq.~(\ref{oneway}) as
\begin{multline}
\frac{\nu_2-\nu_0}{\nu_0}
= -\frac{a^*}{c}T + \frac{1}{2}(1+\beta) \left[1-\frac{a^*}{2c}T \right]
h(t)
\\ -\beta \left[ 1-\frac{a^*}{2c}T \right] h\big( t + (1-\beta)T/2 \big)
-\frac{1}{2}(1-\beta) \left[1-\frac{a^*}{2c}T \right] h(t+T)
\label{Eq:Berrygwshift}\, ,
\end{multline}
where $T$ now denotes the unperturbed {\em two-way} light time and $\nu_2$ 
is
the frequency at reception \cite{Esta:1975,Hellings:1981xc}.  The anomalous
blueshift contributes to the frequency shift by two types of terms. On the 
one
hand it arises proportional to the unperturbed light travel time. If the 
light
travel time is time dependent $T(t)$, as will be the case for LISA, the
influence of the anomalous blueshift arises at the different frequencies
contained in $T(t)$ and at null frequency anyway. On the other hand the
anomalous frequency shift appears as a crossterm with the GW strain. This 
effect
is hence suppressed by the smallness of the GW strain but still several
orders of magnitude larger than terms quadratic in the GW strain.  Both
manifestations of the anomalous blueshift will be investigated in the
following.

\section{Frequency domain analysis \label{frequency}}

In the previous section, we have discussed how the Pioneer anomaly could 
find
its explanation in a blueshift of light and we found the
generic Doppler response function to a plane GW in the
presence of an anomalous blueshift.  This Doppler response function 
describes
the influence of the anomalous blueshift on GW detection by an 
interferometer
arm of LISA.  In the following we analyse this signal
both {\em inside} and {\em outside} the sensitivity band of the LISA.  This 
is
done first through an analysis in the frequency domain \cite{Giamp:1996} and
afterwards in the framework of TDI \cite{Tinto:1999yr,Tinto:2004wu}.

The noise cancelation algorithm in the frequency domain
\cite{Giamp:1996,High} is now considered as obsolete for the LISA
data analysis and has been superseded by TDI, which achieves a far
superior cancellation of laser phase noise in the signal than the
frequency domain algorithm. For our purposes the analysis in the
frequency domain has the considerable advantage that it allows an
intuitive understanding of the influence of the anomalous
blueshift. In this method the magnitude of various contributions
to the signal can be easily compared and have a direct
interpretation in term of phase shifts of an idealised signal.

This convenient interpretation is partially lost in TDI, in which 
combinations
of signals are formed following an algebraic method in order to cancel the
dominant noise source of the interferometer. A more physical picture can in
some part be regained by interpreting the TDI combinations as synthesised
interferometers. Nevertheless an investigation of the impact of an anomalous
blueshift purely in terms of TDI might miss important effects of the 
blueshift,
which might be cancelled by the specific symmetries of the TDI observables. 
On
the other hand the possibilty exists that TDI combinations become 
particularly
sensitive to the blueshift on behalf of their symmetries. Hence it is
important to investigate if TDI remains unimpaired by an anomalous blueshift
and if TDI is capable of detecting a potential anomaly.

Our analysis of the anomalous blueshift in the frequency domain is based on
the method of \cite{Giamp:1996}.
We amend the original discussion by the consideration of additional noise
sources, such as acceleration noise, which were not addressed in
\cite{Giamp:1996} and we update the values of laser and shot noise to match
the current expectations for LISA (cf.\ \cite{LISA,Tinto:1999yr}).

\subsection{The two-way Doppler signals}\label{sect:twow}

For simplicity, we assumed in our analysis that each laser has the same
fundamental frequency $\nu$, whereas in a realistic LISA configuration the
frequency of the lasers may differ to each other by several hundreds of
MHz. As a further simplification we start our discussion by assuming
constant and exactly known (but unequal) lengths of the interferometer arms.
This assumption will be dropped later.

\begin{figure}[ht]
\begin{center}
\includegraphics[width=7cm]{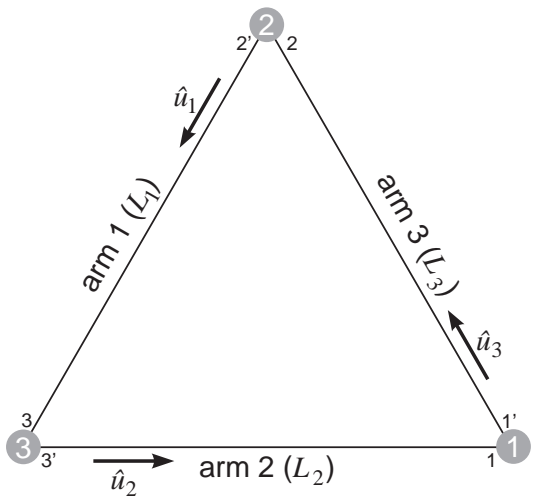}
\end{center}
\caption{Geometry of the LISA formation.} \label{Fig:Unequal}
\end{figure}

The basic interferometer configuration is displayed in the
Fig.~\ref{Fig:Unequal}.  The distances between pairs of spacecraft are $L_1$,
$L_2$ and $L_3$, with $L_i$ corresponding to the interferometer arm opposite
to spacecraft $i$. The optical benches of each spacecraft are labelled by a
number, which corresponds to that of the host spacecraft. An apostrophe allows
to distinguish the two optical benches of the same spacecraft. In addition, a
unit vector $\hat{u}_i$ is asigned to each arm, with $i$ being the label of
the opposite spacecraft.  The orientation of the three unit vectors are such
that $\hat{u}_1 L_1+\hat{u}_2 L_2+\hat{u}_3 L_3=0$.

The phase of the signal received from a distant spacecraft of the
LISA constellation is the sum of the following contributions:
\begin{enumerate}
\item The phase $2\pi \nu l_i(t)$ due to the runtime of the signal, where
  $l_i(t)$ is the one-way light-time for the signal along the $\emph{i}$th
  interferometer arm.  It changes due to the slow relative velocities between
  the spacecraft and on shorter timescales due to GWs.

\item The laser phase noise, $p_i (t)$, is the phase noise of the $i$th 
laser, so
that the phase of the $\emph{i}$th laser is $P_i= 2\pi \nu t+p_i(t)$.

\item The shot noise. Its effect is immediate at the time of reception, so
that the response
of the Doppler measurement at the $\emph{i}$th laser is simply
given by $n_i(t)$.

\item The acceleration noise. The phase variation $\Delta\varphi$ of a
signal depends on
the path length $x$ through
\begin{equation}\label{convert}
\Delta\varphi=\frac{2 \pi \nu}{c} \Delta x \, .
\end{equation}
Therefore, the residual acceleration $\vec{a}_i(t)$ of the optical
bench of the $i$th spacecraft appears in the second derivative of
the phase of the signal. Obviously, the residual accelerations at
the two spacecraft, both at transmission and reception, have to be
taken into account according to the following expression
\begin{equation}
\Delta\varphi_{i}=\frac{2 \pi \nu}{c}\int
\int 
\left[\hat{u}_{j}\cdot\vec{a}_{i}(t)-\hat{u}_{j}\cdot\vec{a}_{k'}(t-l_j)
\right]
dt^2 \label{accn} \, ,
\end{equation}
where $i$ and $k'$ are the end lasers of arm $j$ and
$\Delta\varphi_{i}$ is the phase variation at the photodiode $i$.
Note that this equation considers only the acceleration along the
optical axis and does not take into account a possible turning of
the optical bench.

\item The anomalous blueshift. Using Eq.~(\ref{blue}) and defining
$\alpha^* \equiv - a^*/c$ its contribution to the
phase of the signal is given by
\begin{equation}
\Delta \varphi^*_i(t)=2\pi \alpha^* \int \Delta l_i(t) dt \, .
\end{equation}
\end{enumerate}

By taking into account all of the above contributions, the phase
of the signal sent by the $\emph{k}'$th laser and received at the
$\emph{i}$th reads
\begin{multline}
\varphi_i(t) =2\pi\nu \,\left( t-l_j(t) \right)
+p_{k'}(t-l_j)+n_i(t) \\ + 2\frac{\pi \nu}{c}\int
\int[\hat{u}_{j}\cdot\vec{a}_i(t)-\hat{u}_j\cdot\vec{a}_{k'}(t-l_j)]
dt^2 +2\pi\alpha^*\int l_j(t)dt \, .
\end{multline}
At the reception, the incoming
signal is beaten with the signal $P_i$ of the local laser to give
the beat signal
\begin{equation}
\sin(\varphi_{i}(t))+\sin(P_i(t))=
2\sin \left[ \frac{\varphi_{i}+P_i}{2} \right]
\cos \left[ \frac{\varphi_{i}-P_i}{2} \right]\label{beat} \, .
\end{equation}
The high-frequency sine term is too fast to be read and is not
used in the data analysis. Therefore, on the \emph{j}th arm, the
phase of the beat signal read in the spacecraft photodiode is
given by
\begin{eqnarray}
s_i(t)&=&\varphi_i(t)-P_i(t)\nonumber\\&=&-\,2\pi\nu
l_j(t)+p_{k'}(t-l_j)-p_i(t)+2\pi\alpha^*\int
l_j(t)dt+n_{k'}(t)\nonumber\\
&&+\,2\frac{\pi \nu}{c}\int
\int[\hat{u}_j\cdot\vec{a}_i(t)-\hat{u}_j\cdot\vec{a}_{k'}(t-l_j)]
dt^2
\, ,
\end{eqnarray}
where we have dropped the factor $1/2$ from the argument of the cosine in
Eq.~(\ref{beat}). Furthermore the two lasers on each spacecraft are tied to
each other in phase by the exchange of a two-way reference signal between
them.

The two-way Doppler signal is then formed by the combination of
the phase measurements from two photodiodes on the same arm (cf.\
\cite{Giamp:1996}),
\begin{eqnarray}
z_i(t)&=&s_i(t)+s_{k'}(t-l_j)\nonumber\\
&=&p_i(t-2l_j)-p_i(t)-4\pi\nu l_j(t)+4\pi \alpha^*\int
l_j(t)dt+n_{k'}(t)+n_i(t-l_j)\nonumber\\
&&+2\frac{\pi \nu}{c}\int
\int[\hat{u}_j\cdot\vec{a}_i(t)-2\hat{u}_j\cdot\vec{a}_{k'}(t-l_j)+\hat{u}_j\cdot\vec{a}_i(t-2l_j)]
dt^2\, . \label{z1}
\end{eqnarray}
In order to obtain $z_i(t)$, $s_{k'}(t)$ is sent to the
$\emph{i}$th laser to be beaten with $s_i(t)$. Here, the beat
signal is filtered in order to preserve the GW
contribution, i.\,e.\ by reading the cosine term in the expression of
a beat (analogous to Eq.~(\ref{beat})).

\subsection{Inside the sensitivity band \label{inside}}

In Eq.~(\ref{z1}), the light travel time as a function of reception time
$l_j(t)$ includes both the orbital motion of the spacecraft and the GWs.  We
write explicitly the contribution of the GWs by now considering $l_j(t)$ in as
the nominal arm length in the undisturbed spacetime and adding the disturbance
by the GW as a separate term. Using the Doppler response function,
Eq.~(\ref{oneway}), the effect of a GW, transverse to the LISA plane, i.\,e.\ 
$\beta \equiv 0$, and with appropriate polarisation, on the two-way Doppler
signal is given by
    \begin{equation} \label{zh}
    \frac{\Delta \nu}{\nu}=
    \frac{1}{2}\epsilon(1+\alpha^*l_j)[h(t)-h(t-2l_j)] \, ,
    \end{equation}
    where $\Delta \nu$ is the difference between the frequency of the signal
    sent and received at the central spacecraft and $h$ is the GW strain 
amplitude. The $\epsilon$ can take any value between -1 and 1,
    depending on the orientation of the arm with respect to the polarisation
    of the GW. Particularly, for an angle of $60^\circ$ between the arms of 
LISA,
    one can have $\epsilon = 1$ for one
    arm and $\epsilon = -1/2$ for the other (cf.\ \cite{LISA} p.~102 for the
    general expressions). The GW adds a contribution to the signal,
    Eq.~(\ref{z1}),
\begin{equation}
\Delta \varphi_{\text{gw}}= \epsilon \, \pi \nu
\int(1+\alpha^*)[h(t)-h(t-2l_j)]dt\, .
\end{equation}
In order to estimate the importance of each term in Eq.~(\ref{z1})
and (\ref{zh}), it is useful to compute the power spectral density
of $z_i(t)$. To begin, we restrict our study to the sensitivity
band of LISA, i.\,e.\ from $10^{-4}\,{\rm Hz}$ to $1\,{\rm Hz}$. In a first
estimate we can drop the two terms, $-4\pi\nu l_j(t) +4\pi
\alpha^*\int l_j(t)dt$ because the orbital motion has little
impact at the frequencies of the sensitivity band. To compute the
power spectral density, we consider the Fourier
transform of $z_i(t)$,
\begin{multline}\label{ZZZ}
z_i(f)=p_i(f)[e^{4 \pi i f l_j}-1]+n_i(f)[1+e^{2\pi i f
l_j}] \\ +\nu a_i(f)\frac{[e^{4 \pi i f l_j}+2e^{2\pi i f
l_j}+1]}{2\pi c f^2}
+\epsilon \nu h(f)\frac{[e^{4\pi i fl_j}-1]}{2if} +\epsilon \nu
\alpha^*l_j \, h(f)\frac{[e^{4\pi i fl_j}-1]}{2if} \, ,
\end{multline}
where we have assumed that the shot noise and acceleration spectra for both
optical benches are the same, $n_i(f)=n_{k'}(f)$ and $a_i(f)=a_{k'}(f)$.
Furthermore we have assumed the maximum value for the direction cosine between
$\vec{a}_{i,k'}$ and $\hat{u}_j$.  The Fourier transform, Eq.~(\ref{ZZZ}),
supposes that the observing time is infinite. In practice, LISA is expected to
operate in data-taking intervals of $T\sim 10000\,{rm s}$ and thus,
Eq.~(\ref{ZZZ}) only gives an estimate of the true spectrum. We will return to
the effect of finite observation time below.

From Eq.~(\ref{ZZZ}) it can be read off immediately that the effect of the
anomalous blueshift would be undetectable. The blueshift enters the spectrum
folded with the GW strain $h$. Hence its effect will be 15
orders of magnitude below the GW signal. This corresponds to a
spectral power at least 10 orders of magnitude below the secondary noises,
shot noise and acceleration noise
(cf.\ \cite{LISA,Tinto:1999yr} for the estimated noise spectra for LISA).
Currently no procedure exists
to cancel the shot noise in the LISA signal.  Hence
the anomalous blueshift would be overwhelmed by the
secondary noises and remain unnoticed.

\begin{figure}[ht]
\begin{minipage}[t]{0.48\textwidth}
\centering
\includegraphics[width=\textwidth]{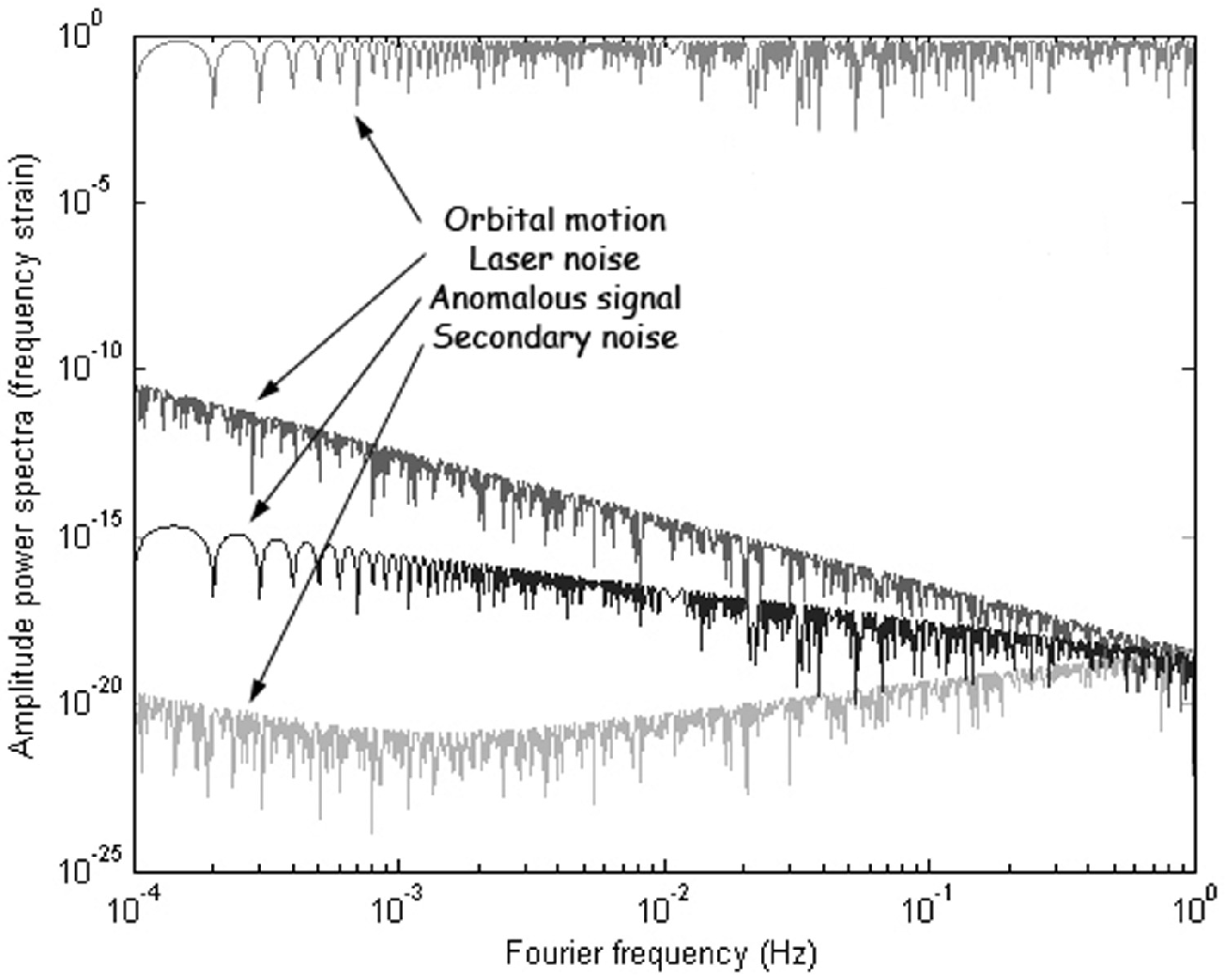}
\caption{Amplitude power spectra contributing to the two-way
Doppler signal in the sensitivity band.}
\label{fig:tidju}
\end{minipage}
\hfill
\begin{minipage}[t]{0.48\textwidth}
\centering
\includegraphics[width=\textwidth]{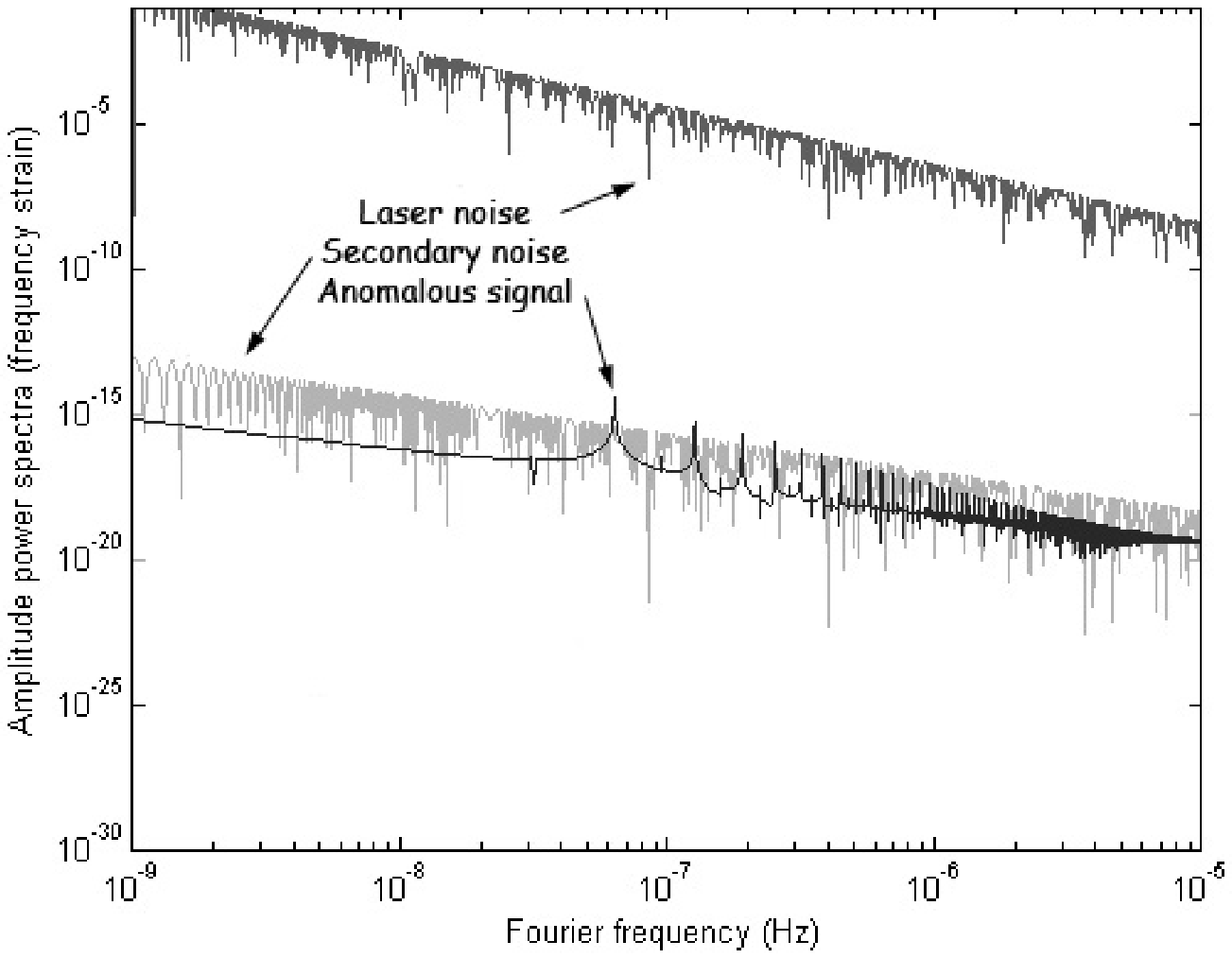}
\caption{Amplitude power spectra contributing to the two-way Doppler signal 
outside the sensitivity
band.} \label{fig:out}
\end{minipage}
\end{figure}

This conclusion has however to be reconsidered taking into account that the
data-taking periods of LISA are limited in length. This
leads to the leakage of spectral power to other
frequencies. In particular, the low-frequency terms neglected in
the two-way signal $z_i(t)$, Eq.~(\ref{z1}), have now to be
addressed.
For typical integration times, $T \sim 10000\,{\rm s}$ (cf.\
\cite{Tinto:1999yr}), the arm length rate of change is nearly
constant. Its magnitude depends on the position of the spacecraft
along its orbit \cite{Folkner:1997hn}. The relative velocity,
$\vec{v}$ can reach up to $13\,{\rm m/s}$.\footnote{See however
\cite{Nayak:2005kb} for recent suggestion of a modified orbit,
which could reduce the relative velocity between the spacecraft by
a factor of six.} In the approximation of constant relative
velocity the Fourier transform of $l_j(t) = l_{j,0} + (v_j/c) t$ is
given by
\begin{multline}
\widetilde{F.T}.[l_j(t)]=\int_{0}^{T} \left(l_{j,0} +
\frac{v_j}{c}t\right)e^{2\pi
ift}dt\\
=\frac{v_j}c \, e^{i\pi Tf}\frac{\pi Tf\cos(\pi T f)-\sin(\pi T
f)}{2\pi^2if^2}+\left[l_{j,0} + \frac{v_j}{2c} T \right]e^{i\pi T
f}\frac{\sin(\pi T f)}{\pi f}\, ,\label{Eq:leak}
\end{multline}
where $cl_{j,0} \approx 5\times10^9$\,m is the initial
light time between the spacecraft. In Eq.~(\ref{Eq:leak}) the constant term
due to the arm length, $l_{j,0}$, is dominant. This term will
remain present even after the application of the laser noise
cancellation algorithm in the frequency domain (cf.\
\cite{Giamp:1996}).
Using Eq.~(\ref{Eq:leak}) the Fourier transform of the two-way Doppler 
signal
becomes
\begin{multline}\label{Eq:complete2ways}
z_i(f)= p_i(f) \left[e^{4 \pi i f
l_j}-1\right]+n_i(f)\left[1+e^{2\pi i f l_j}\right]+ \\
\nu a_i(f)\left[e^{4 \pi i f l_j}+2e^{2 \pi i f l_j}+1\right]
+4\pi \nu l_j(f)-\frac{2\alpha^*}{if}l_j(f)\, .
\end{multline}
In this expression, we have dropped the contribution of GWs because above it
was found irrelevant for the discussion of the blueshift (See
\cite{Giamp:1996} for a discussion of the GW signal in terms of the 
frequency
domain algorithm). The contributions to the amplitude power spectrum
corresponding to Eq.~(\ref{Eq:complete2ways}) are shown in 
Fig.~\ref{fig:tidju}. The signal of the anomalous blueshift is higher than 
the
secondary noise sources but below the laser phase noise. The nominal term 
from
$l_j$, i.\,e.\ the orbital motion is much higher than the laser noise. Hence
this term would have to be removed by a preprocessing method before the 
laser
noise cancellation algorithm could be applied (see below). The signal of the
anomalous blueshift is below the laser phase noise but above the secondary
noise sources.

The further processing of the signal is distinct for the search for
GWs and for the search for an anomalous blueshift.  In the
search for GWs the spectral leakage is unwanted.  Hence a
suitable approach is pre-multiplying the time-domain data sets by a window
function before taking Fourier transform \cite{Tinto:1999yr}. With this
preprocessing, the laser noise cancellation can be performed and GWs could 
be detected.  On the other hand, in the search for an anomalous
blueshift the spectral leakage has to be preserved. However before the
cancellation of the laser noise could be attempted, one has to 
generate a
signal in which the laser noise is the dominant disturbance. Hence the 
nominal
orbital term needs to be removed from the signal. The natural approach to 
this
task would be to pre-process the data with information on the orbits 
acquired
from a different source, e.\,g.\ ground tracking of the spacecraft.

The crucial question is how accurately we would need to determine
the arm length of LISA to sufficiently remove the nominal orbital
motion term. According to Fig.~\ref{fig:tidju}, a cancellation of
about $10^{15}$ orders of magnitude would have to be performed. We
suppose that the arm length $l_j$ is known up to a factor $k$,
i.\,e.\ the real arm length $l_{j}$ differs from the assumed arm
length $l_{j,\text{measured}}$ by the length $k l_j$,
$l_{j,\text{measured}} = (1+k) l_j$.  Then the Fourier transform
of the one-way Doppler signal becomes
\begin{multline}\label{arghhhh}
z_i(f) = p_i(f) \left[e^{4 \pi i f l_j}-1 \right]
+n_i(f) \left[1+e^{2\pi i f l_j} \right] \\
+a_i(f) \left[e^{4 \pi i f l_j}+2e^{2 \pi i f l_j}+1 \right]
+4(1+k)\pi \nu l_j(f)-(1+k)\frac{2\alpha^*}{if}l_j(f)\, .
\end{multline}
Therefore, after the removal of the nominal orbital motion term
from the knowledge of $l_{j,\text{measured}}$, the term $4k\pi \nu
l_j(f)$ remains which has to be sufficiently low to detect the
anomaly. However at $10^{-4}$\,Hz, this requirement corresponds to
a knowledge of the arm length of about $10^{-15}\times l_j=
5\times10^{-6}\,{\rm m}$, which is far beyond the experimental
capabilities of LISA.

In conclusion an anomalous blueshift of the magnitude of the
Pioneer anomaly would remain undetected in the sensitivity band of
LISA. Furthermore it would not affect LISA's capability to observe
GWs. The blueshift remains unimportant because it
is peaked around zero frequency and thus far away from the sensitivity band 
of
LISA. However it is still worth considering the potential impact of the
anomaly at frequencies below the sensitivity
band of LISA where the effect becomes much larger.

\subsection{Outside the sensitivity band \label{outside}}

For long integration times, the arm-length rate of change cannot
be treated as constant anymore. In the following we use the
simplified analytical model of the LISA orbits described in
\cite{Dhurandhar:2004rv}, in which only the Kepler
problem for each spacecraft is considered. Computing the
corresponding power spectrum, one can plot the two-way Doppler
signal, outside the sensitivity band of LISA.

Unfortunately, the noise spectra for LISA at frequencies below the 
measurement
band have not yet been fully investigated (cf.\ \cite{Bender:2003uv}).  For
our purposes we use an extrapolation of the noise spectra obtained in
\cite{Bender:2003uv}. The acceleration noises might become considerably 
higher
if a suspension mode for low frequencies would be implemented along the
optical axes of the interferometer. However the results of
\cite{Industry:2000} indicate that the best performance of LISA is obtained 
if
the drag-free mode along the sensitivity axes is maintained also for
low-frequencies. Hence an extrapolation of the noise spectra given in
\cite{Bender:2003uv} should give a reasonable impression of the actual
performance to expect from LISA at low frequencies.

The result is displayed in Fig.~\ref{fig:out}.
The term due to the anomalous blueshift is of the same order of
magnitude as the secondary noise but remains still below the laser
phase noise. However, for the integration time required to reach
these frequencies, the laser noise cancellation algorithm can no
longer be implemented because the arm length changes from the
orbital motion of LISA are so big that the algorithm becomes
ineffective \cite{Giamp:1996,High}. Thus the presence of an
anomalous blueshift cannot be revealed at low frequencies either.

To summarise, the anomalous blueshift would have an amplitude several orders
of magnitude higher than the weakest GWs detectable, on the
LISA's arms. However, this ``large'' impact of the anomaly comes from the
constant part of the arm lengths. Hence it is located at null Fourier
frequency while relevant GWs for LISA are expected at Fourier
frequencies between $10^{-4}\,{\rm}$ and $1\,{\rm Hz}$. In the sensitivity
band of LISA, the effect of the anomaly is well below all the instrumental
noises and hence is neither detectable nor does it have an impact on the
GW detection.  With a finite observation time, power of the
constant contribution of the anomaly can leak in the sensitivity band.  The
analysis of the spectral leakage of the anomalous blueshift would however
require a knowledge of the arm length that would have to be far more precise
than it is achievable.  Below the sensitivity band, we found that the
contribution of the anomalous blueshift should be just above the secondary
noise sources but still below the laser phase noise.  On these timescales, 
the
arm lengths change much more than it is allowed to remove efficiently the
laser phase noise.  Therefore, we can conclude that the Pioneer anomaly has 
no
impact on the GW detection and cannot be detected with the
frequency domain method.

\section{Time delay interferometry}\label{time}

Time delay interferometry (TDI) is a noise cancellation method for unequal 
arm
interferometers that is performed in the time domain
\cite{Tinto:1999yr,Tinto:2004wu,Estabrook:2000ef,Tinto:2002de} (see also the
contribution by M.\ Tinto in this volume). The basic principle of TDI 
consists
in combining appropriate one-way Doppler signals in order to remove the 
laser
phase noise (actually, it also cancels the acceleration noise of the optical
benches).  Whereas TDI had originally been developed as a purely algebraic
method its data combinations have a physical interpretation as virtual
measurements of a synthesised interferometer \cite{Vallisneri:2005ji}.  The
major question to be addressed here is how the Pioneer anomaly affects the 
TDI
combinations. Since the frequency domain study showed that the effect of the
anomalous blueshift is negligible when folded with the GW
strain, we can restrict our attention on the anomalous blueshift occurring 
at
low frequencies.

\subsection{Linear Data combinations}\label{combi}

In principle there is an unlimited number of TDI observables corresponding 
to
more and more complicated synthesised interferometers.
For applications to LISA
the number of beams to combine is usually limited to eight in the limit of a
static interferometer. For this maximum number of beams
there are ten linear combinations, which cancel the laser noises
from all the spacecraft.

These TDI combinations cancel the laser phase noise of an interferometer at
rest with unequal but constant arm lengths and are commonly dubbed {\em 
first
  generation TDI}.  For the nominal operation mode of LISA, the unequal-arm
Michelson interferometer, three independent possible combinations exist 
which are
called $X$,$Y$ and $Z$. In the following we only consider the $X$ 
combination,
\begin{multline}\label{Eq:x}
X = 
y_{32,322}-y_{23,233}+y_{31,22}-y_{21,33}+y_{23,2}-y_{32,3}+y_{21}-y_{31}
 \\
+ \frac{1}{2}(-z_{21,2233}+z_{21,33}+z_{21,22}-z_{21})
+\frac{1}{2}(+z_{31,2233}-z_{31,33}-z_{31,22}+z_{31})\, .
\end{multline}
The Doppler data to be analysed are now called $y_{ij}= \Delta
  \nu / \nu$, where $\Delta \nu$ is the frequency deviation from the centre
frequency $\nu$.  The subscripts label the transmitting and receiving
spacecraft. The convention is that $y_{ij}$ is the beam transmitted from
spacecraft $i$ and received at spacecraft $j$. Internal metrology signals to
correct for optical bench motions are denoted by $z_{ij}\,$, with the same
labelling convention. These will however play no role in our considerations
because their travel times are too short to show an appreciable anomalous
blueshift. They are hence omitted from now on. Delay of laser data streams
is indicated by commas in the subscripts: $y_{31,23} =
y_{31}(t-l_2-l_3)=y_{31,32}\,$, etc.\ ($l_i$ is the light-time on the
$\emph{i}$-arm).

The $Y$ and $Z$ combinations are obtained from the $X$ combination
by cyclic relabelling of the spacecraft. Hence our results hold
for all three of the unequal-arm Michelson combinations.  It is
easy to verified by direct substitution of the laser noise
contribution, that the combination, Eq.~(\ref{Eq:x}), does not
contain any laser noise. In the unequal-arm Michelson
combinations, each one-way signal occurs twice, at two different
times; one term is added and the other subtracted. As a
consequence, the Pioneer anomaly component, given by
Eq.~(\ref{blue}),
\begin{equation}\label{Eq:tdianomaly}
y^{*}_{ij}=-\frac{a^*}{c}l_i(t)= \alpha^* l_i(t) \, ,
\end{equation}
disappears. Even the spectral leakage of the data has no impact because the
terms, which contain the anomaly are all cancelled {\em exactly}. The same
property holds for the other combinations of the data present in the
literature (see \cite{Tinto:2004wu} for a description of the other
combinations).

In the combinations called $\alpha$, $\beta$, $\gamma$ and $\zeta$, which
represent synthetic Sagnac interferometers, the contribution of the anomaly
would not be cancelled if the frequency shift would depend on the direction 
of
the light beam with respect to the Sun. To obtain such a direction
dependent anomaly, that does not decay significantly over tens of AUs, one 
would
however have to resort to exotic ideas like a topological defect located in
the Sun.  Such a model is hard to envisage and no such effect has been
suggested as an explanation of the Pioneer anomaly. Hence we do not
further consider this possibility.

\subsection{Effect of the orbital motion on Time Delay
Interferometry}\label{sect:TDI2}

The first generation TDI observables, as presented above, have been 
formulated
in the limit, that LISA is fixed in space. However, each year, LISA will
accomplish a complete rotation around its centre and the symmetries provided
by a fixed interferometer will be broken. Because of this loss of symmetry, the
contribution of the anomalous blueshift, arising on each arm, would not be
cancelled completely anymore in the TDI combinations. Moreover, the laser
phase noise does not cancel exactly, either. More complicated TDI 
combinations
have been developed to overcome this problem
\cite{Cornish:2003tz,Shaddock:2003dj,Tinto:2003vj}. In addition to the
rotation, there occurs a flexing of the arms of the detector, which is 
caused
by the orbital motion and the perturbations of the planets. The interaction 
of
these two effects with the anomaly is considered in the following.

\subsubsection{The effects of rigid rotation}

For the discussion of rigid rotation, a more subtle notation for
the Doppler signals is required because the light travel times
will now depend on the direction of the signal with respect to the
rotation.  While in the previous section $L_3$ was the length of
the arm between the first and the second spacecraft, we denote now
by $L_{12}$ the length travelled by the signal sent from
spacecraft~1 and received at spacecraft~2.
The length travelled by the signal sent from spacecraft~2 and
received at spacecraft~1 is called $L_{21}$.
\begin{figure}[ht]
\begin{center}
\includegraphics[width=7cm]{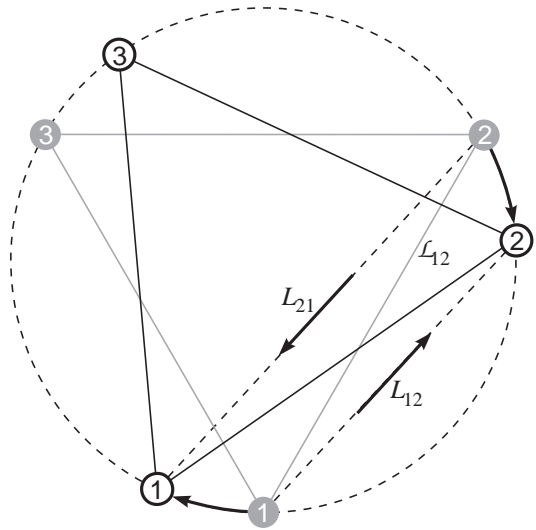}
\end{center}
\caption{The rotation of the interferometer breaks the direction
symmetry in the arm lengths.} \label{fig:tdirotate}
\end{figure}
As illustrated in Fig.~\ref{fig:tdirotate} the interferometer is rotating in
the clockwise direction if viewed from the celestial pole. The spacecraft move
while the signals are travelling along the arms. If we define the length of
the arm between spacecraft 1 and 2 to be $\mathcal{L}_{12}$ in the limit of no
rotation, then the actual distance travelled by the signal from spacecraft 1
to spacecraft 2 will be $L_{12}<\mathcal{L}_{12}$. In the same manner, the
signal from spacecraft 2 to spacecraft 1 will have to reach spacecraft 1 in
its motion and will therefore travel a distance $L_{21}>\mathcal{L}_{12}$.
Hence also the magnitude of an anomalous blueshift on an arm would depend on
the direction, in which the signal has travelled. Then, if the signals, which
have travelled on the same arm but in opposite direction, are subtracted, a
residual contribution of the anomalous blueshift would remain.

The direction dependence of the light travel times has different effects on
the induvidual TDI combinations.  For the unequal-arm length interferometric
combinations $X(t)$, $Y(t)$ and $Z(t)$, the contribution of the anomaly still
cancels exactly. Indeed, if we take the $X(t)$ combination (the reasoning is
the same for $Y(t)$ and $Z(t)$), we see that the one-way signals appear twice
for a given direction with opposite signs in the combination but that they are
delayed by different times. For a rigid rotation the relation
$L_{ij}(t+\tau)=L_{ij}$ holds and the contribution of the anomaly is
cancelled. The ($P,\,Q,\,R$), ($E,\,F,\,G$) and ($U,\,V,\,W$) have the same
property so that one reaches the same conclusion for these combinations.

For the Sagnac combinations ($\alpha$, $\beta$, $\gamma$ and $\zeta$), the
structure of the signal is different (cf.\ \cite{Tinto:2004wu}). In these
observables two signals from each arm, running in opposite direction are
combined. For instance the signal $\zeta$ reads,
\begin{equation}\label{Eq:zeta2}
\zeta =y_{32}-y_{23,3}+y_{13,3}-y_{31,1}+y_{21,1}-y_{12,2}\, .
\end{equation}
Hence the effect of the anomaly is not totally removed. The anomalous
component in Eq.~(\ref{Eq:zeta2}) reads
\begin{equation}\label{Eq:zeta3}
\zeta^* =\alpha^* (l_{12}-l_{13}+l_{23}-l_{21}+l_{31}-l_{32}) \, .
\end{equation}
The terms of this equation can be grouped into two parts: $\Delta l_- \equiv
l_{12}+l_{23}+l_{31}$, which is the total time around the interferometer in
the counterclockwise direction and $\Delta l_+ \equiv l_{13}+l_{21}+l_{32}$,
which is the total time in the clockwise direction. Even for a perfectly 
rigid
triangle, the times of flight are not equal.
Since the LISA constellation rotates in clockwise direction, we always have
$\Delta l_{-} < \Delta l_{+}$. The corresponding Sagnac time shift is given
by, cf.\ \cite{Cornish:2003tz},
\begin{equation}\label{Eq:sagnaceffect}
\Delta l_{-}-\Delta l_{+}=\Delta l_{\text{Sagnac}}=\frac{4A\Omega}{c^2}
\approx \frac{2\pi \sqrt{3}L^2}{c^2T} \, .
\end{equation}
Here $\Omega$ is the angular velocity of the rotation,
$A$ is the area enclosed by the light path, $T$ is the period of
rotation and $L$ is a typical arm length.
For the LISA orbit
($T =1$\,year and $L=5\times10^{9}$\,m), the Sagnac effect has the magnitude
$\Delta l_{\text{Sagnac}} =- 10^{-4}$\,s. Therefore, the residual
effect of the anomalous blueshift on the combination $\zeta$ would be
\begin{equation}\label{Eq:zeta5}
\zeta^* =\alpha^* (\Delta l_{-}-\Delta l_{+})\simeq
3\times10^{-22}\, .
\end{equation}
The same result is obtained for the other Sagnac combinations $\alpha$,
$\beta$, $\gamma$. The effect of the anomalous blueshift is to add a 
constant
frequency shift in the Sagnac combinations. The amplitude of this additional
Doppler shift would be comparable to the weakest GWs detectable by LISA.
However, the optimal sensitivity of LISA occurs in a Fourier frequency range
far from the zero Fourier frequency, where the constant residual contribution
of the anomaly has its impact. Hence, again the effect would not be
detectable.

As mentioned above the rigid rotation induces a Sagnac effect on the noises 
as
well. In order to maintain noise cancellation up to linear order in the
rotational velocity for all observables {\em modified TDI} has
been introduced, in which each Doppler signal from a specific arm enters 
twice
travelling in {\em the same} direction. In modified TDI the anomalous
blueshift is cancelled at linear order in the rotational velocity
and the effect of the anomaly would become even smaller.

\subsubsection{The effects of flexing}

As we have seen in Section \ref{outside}, the arm lengths of LISA don't 
remain
constant due to the orbital motion and the
perturbations of the orbits by the planets. Unlike the rigid rotation, the
flexing of the arms does not preserve the continuous symmetry,
$L_{ij}(t+\tau)=L_{ij}(t)$.
For example, the $X$ combination, Eq.~(\ref{Eq:x}), becomes for time varying
runtimes (cf.\ \cite{Cornish:2003tz})
\begin{eqnarray}\label{Eq:x3}
X&=&y_{12}\left[t-l_{31}-l^{(1)}_{13}-l^{(2)}_{21}\right]-y_{13}\left[t-l_{21}-l^{(1)}_{12}-l^{(2)}_{31}\right]+y_{21}\left[t-l_{31}-l^{(1)}_{13}\right]\nonumber\\
&&-y_{31}\left[t-l_{21}-l^{(1)}_{12}\right]+y_{13}(t-l_{31})-y_{12}(t-l_{21})+y_{31}(t)-y_{21}(t)
\, ,
\end{eqnarray}
where $l_{21}\equiv l_{21}$(t), $l_{31}\equiv l_{31}$(t),
$l^{(1)}_{12}\equiv l_{12}(t-l_{21})$, $l^{(1)}_{13}\equiv 
l_{13}(t-l_{31})$,
$l^{(2)}_{21}\equiv l_{21}(t-l_{31}-l^{(1)}_{13})$ and
$l^{(2)}_{31}\equiv l_{31}(t-l_{21}-l^{(1)}_{12})$.
Using Eq.~(\ref{Eq:tdianomaly}), the contribution of
the anomalous blueshift in the $X$ combination is given by
\begin{eqnarray}\label{Eq:x4}
X^*&=&\alpha^*\left[l_{12}\left(t-l_{31}-l^{(1)}_{13}-l^{(2)}_{21}\right)-l_{13}\left(t-l_{21}-l^{(1)}_{12}-l^{(2)}_{31}\right)+l_{21}\left(t-l_{31}-l^{(1)}_{13}\right)\right]\nonumber\\
&& +\,\, 
\alpha^*\left[-l_{31}\left(t-l_{21}-l^{(1)}_{12}\right)+l_{13}\left(t-l_{31}\right)-l_{12}(t-l_{21})+l_{31}(t)-l_{21}(t)\right]\\
&=&\alpha^*\left[l^{(3)}_{12}-l^{(3)}_{13}+l^{(2)}_{21}-l^{(2)}_{31}+l^{(1)}_{13}-l^{(1)}_{12}+l_{31}-l_{21}\right] 
\nonumber
\, ,
\end{eqnarray}
where $l^{(3)}_{12}\equiv l_{12}(t-l_{31}-l^{(1)}_{13}-l^{(2)}_{21})$
and $l^{(3)}_{13}\equiv l_{13}(t-l_{21}-l^{(1)}_{12}-l^{(2)}_{31})$.
The arm lengths can be estimated by their first order changes
$l^{(n)}_{ij}=l_{ij}(t)-nV_{ij}l$, where $V_{ij}$ is the rate of change of 
the
arm's light travel time in seconds per second and $l$ is a typical one-way
light-time \cite{Cornish:2003tz}. Then we find from Eq.~(\ref{Eq:x4})
\begin{equation}
X^*=4\alpha^*(V_{13}-V_{12})l=6 \times10^{-24} \, ,
\end{equation}
where, in accordance with \cite{Cornish:2003tz}, we have used $V_{13}-V_{12}
\lesssim 10 ({\rm m/s})/c$. This relation determines the maximum effect of the
blueshift due to the flexing induced by the orbital motion. The Doppler shift
stays below the optimal sensitivity of LISA which is about $10^{-23}$.
Results at the same order of magnitude are obtained for all TDI combinations.
Hence we conclude that the effect of the anomalous blueshift due the flexing
of the interferometer arms will not be detectable.

The flexing of the arms inhibits laser noise cancellation to first order in
the velocities for first generation TDI or modified TDI. To achieve the 
noise
cancellation up to first order, inclusive, another set of observables has 
been
designed. The so-called {\em second generation TDI} achieves the 
cancellation
by having not only a signal for each arm entering twice in the same 
direction
but also having each term linear in the change rate of the length of an arm
entering twice. Hence second generation TDI also cancels the anomalous
blueshift from terms linear in the change rate of the arms.

We found that TDI observables, especially those of second generation TDI, 
are
particularly insensitive to the anomalous blueshift. This result has a 
simple
geometric justification. The anomalous blueshift of sizable magnitude arises
proportional to arm-length differences. However TDI is based on combining
signals in a way, that yields overall light-times of zero (cf.\
\cite{Vallisneri:2005ji}). Hence the very principle of the TDI algorithm 
leads
to an automatic cancellation of the anomalous blueshift in the signal.

\section{Summary and conclusions \label{conc}}

The Pioneer anomaly is attracting a growing interest in the scientific
community. Hence a verification of the effect beyond the Pioneer data would be
highly desirable. Here we studied in which way the LISA mission could
contribute to a test of the Pioneer anomaly.  Due to its Earth-like orbit 
LISA
would most likely not experience an anomalous force since this would require 
a
strong violation of the weak equivalence principle. On the other hand, if 
one
interprets the Pioneer anomaly as an anomalous blueshift of light this 
effect
would affect also the LISA interferometer.

Several models in the literature consider the Pioneer anomaly as a homogeneous
and isotropic blueshift originating from the cosmic expansion through various
mechanisms \cite{Rosales:1998mj}-\cite{Ranada:2004mf}. All of these distinct models lead to a common
Doppler response function for LISA up to linear order in the anomalous
blueshift and in the GW strain. We derived this Doppler response function as
an extension of the well known two-point response to GWs. We found that the
blueshift arises on the one hand as a cross term with the GW signal and on the
other hand as a low-frequency bias depending on the interferometer arm length.
The cross term with the GW signal is much larger than a possible second-order
GW term but still too small to be detectable by LISA.

The low-frequency term was found to induce a relative frequency shift of
$10^{-16}$, which is several orders of magnitude larger than the weakest
measurable GW strain of $10^{-23}$. The implications of this 
number
are however not immediate because the anomalous blueshift arises at zero
frequency whereas the LISA sensitivity lies between $10^{-4}\,{\rm Hz}$ and
1\,Hz.

Consequently we investigated the power spectral density of the anomalous
frequency shift, that arises from the orbital motion of the LISA satellites.
We considered both, short times and timescales, which comprise a considerable
fraction of the orbital period of LISA. These results were compared with the
noise spectra of LISA.  Unfortunately, due to its low-frequency nature the
anomalous blueshift is always overwhelmed by some noise source of the LISA
interferometer. Hence an anomalous blueshift would remain undetectable.

This conclusion is then reconsidered in the framework of TDI, the current
baseline method for laser phase noise cancellation in LISA. The dedicated
discussion of TDI is necessary because the complicated signals synthesised in
TDI could produce a by-chance amplification of a homogeneous isotropic signal.
Our results show on the contrary that TDI cancels the blueshift in all data
combinations to a high degree due to the inherent symmetries of the TDI
observables.  Only from the rotation and flexion of the interferometer, a
residual contribution of the anomaly would arise. This effect would however be
below the detection threshold of LISA.  Hence an anomalous blueshift will not
be recognisable in TDI and will not degrade the performance of TDI for the
detection of GWs either.

In the present study we have focused our attention on models of the anomaly,
that predict a homogeneous and isotropic blueshift. At first glance this seems
a bit restrictive, since also models, in which the Pioneer anomaly arises from
a central force, can lead to a considerable blueshift of light. An example is
the model of \cite{Jaekel:2005qe}, which introduces separate
momentum-dependent gravitational couplings for the scalar and the conformal
sector of the Einstein equations. In this model about half of the Pioneer
anomaly is due an anomalous blueshift. However the blueshift induced into the
LISA signal by a central force would be proportional to the difference in
light-travel time between the way back and forth in a two-way signal, whereas
a homogeneous and isotropic blueshift is proportional to the two-way light
travel time. Thus the blueshift from a central force, that is supposed to
explain the Pioneer anomaly, would in general have much less effect on the
LISA signal than a homogeneous and isotropic blueshift.

In conclusion, LISA cannot be used to test the Pioneer anomaly and one will
have to look for other options to verify if the Pioneer anomaly could be a
novel physical effect. Considering the blueshift interpretation of the
anomaly, missions for a test of general relativity by interferometry like
LATOR \cite{Turyshev:2005ux} (see also the contribution by S.~Turyshev in this
volume) or ASTROD \cite{Ni:2002cd} (see also the contribution by Wei-Tou Ni in
this volume) might be sensitive to this effect. However also these missions
would face the problem that the anomalous blueshift becomes significant only
at low frequencies, i.\,e.\ for large changes of the light travel time in the
interferometer.

More promising -- and probably mandatory if the Pioneer anomaly represents a
force and not a blueshift -- would be a test in the outer Solar system by
radio-tracking of a deep space vehicle with very well know onboard 
systematics
\cite{Dittus:2005re}.  Preferably this would be a dedicated mission to 
explore
the anomaly although a planetary exploration spacecraft, which has been
designed with the secondary goal to test the Pioneer anomaly could already
gain considerable insights \cite{Izzo:2005dz}.\footnote{See
  \cite{Rathke:2004gv,Bondo} for an example implementation of a Pioneer 
anomaly
  test on a Pluto exploration mission.} The analysis of the full archive of
Pioneer~10 and 11 Doppler data, that is currently being initiated, might
further help to identify mission scenarios that are especially suited for a
test of the anomaly.

\section*{Acknowledgements}

The authors would like to thank the organisers of the 359th 
WE-Heraeus-Seminar
``Lasers, Clocks and Drag-Free: Key Technologies for Future High Precision 
Test
of General Relativity'' for the realisation of this stimulating conference.
The authors acknowledge support by the Advanced Concepts Team of the 
European
Space Agency where the major part of this work was carried out. This work 
has
much benefitted for discussions with O.\ Jennrich and D.\ Izzo.

\bibliographystyle{unsrt}

\end{document}